\documentclass{article}
\usepackage{arxiv}
\usepackage{moreverb}
\usepackage[utf8]{inputenc} % allow utf-8 input
\usepackage[T1]{fontenc}    % use 8-bit T1 fonts
\usepackage{hyperref}       % hyperlinks
\usepackage{url}            % simple URL typesetting
\usepackage{booktabs}       % professional-quality tables
\usepackage{amsfonts}       % blackboard math symbols
\usepackage{nicefrac}       % compact symbols for 1/2, etc.
\usepackage{microtype}      % microtypography
\usepackage{lipsum}		% Can be removed after putting your text content
\usepackage{graphicx}
\usepackage{natbib}
\usepackage{listings}
\usepackage{doi}

\title{ClusterBuilder - A DSL to Deploy a Parallel Application over a Workstation Cluster}

\date{}

\author{ \href{https://orcid.org/0000-0002-1261-9519}{\includegraphics[scale=0.06]{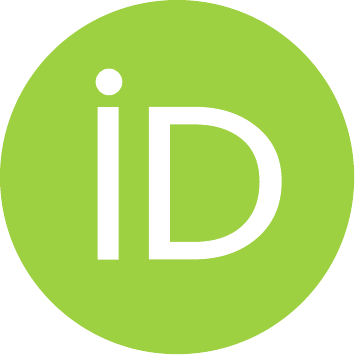}\hspace{1mm}Jon M.~Kerridge} \\
    School of Computing\\
    Edinburgh Napier University\\
    Merchiston Campus\\
    Edinburgh EH10 5DT\\
    \texttt{j.kerridge@napier.ac.uk}}

\renewcommand{\shorttitle}{Cluster Builder DSL}
\renewcommand{\headeright}{Jon Kerridge}
\renewcommand{\undertitle}{}

\begin{document}
\maketitle

\begin{abstract}
Many organisations have a large network of connected computers, which at times may be idle.  These could be used to run larger data processing problems were it not for the difficulty of organising and managing the deployment of such applications.  ClusterBuilder is designed to make this task much simpler.  ClusterBuilder uses its own Domain Specific Language (DSL) to describe the processing required that removes the need for a deep understanding of parallel programming techniques.  The application uses extant sequential data objects which are then invoked in a parallel manner.  ClusterBuilder uses robust software components and the created architecture is proved to be correct and free from deadlock and livelock.  The performance of the system is demonstrated using the Mandelbrot set, which is executed on both a single multi-core processor and a cluster of workstations.  It is shown that the cluster-based system has better performance characteristics than a multi-core processor solution.
\end{abstract}


\section{Introduction}\label{sec1}
The goal for ClusterBuilder was to create a system that enables the deployment of a parallel solution to a problem with few changes to an existing sequential Java code. The solution should exploit parallel techniques and not just run the same sequential application many times on the available workstations.  The aim was to produce a Domain Specific Language (DSL)\citep{domainspecificlanguage} that captured the essence of the required solution and left the detail to a builder application that created the required parallel coding.  The user does not need to be aware of the detailed architectural, network structure and parallel processing requirements.

\subsection{Requirements}\label{sec1.1}
The requirements that guided the design of the DSL and the subsequent deployment of the application were:
\begin{enumerate}
\item{use commodity workstations and operating systems connected using an ethernet}
\item{use extant techniques to transfer files around the network and invoke an application}
\item{minimise the amount of code that needs to be distributed by the user}
\item{define and build application network interconnections with no user intervention}
\item{produce a solution that is provably free from deadlock and livelock}
\item{require no detailed knowledge of the workstation interconnection network}
\item{collect detailed timing data of node usage including load and run-time separately}
\end{enumerate}
Requirement 1 was predicated by wanting to use an organisation’s existing network of workstations without requiring the adoption of new software tools.  Requirement 2 was simply a reflection that Windows provides the Remote Desktop Connection\citep{MSremoteaccess} tool and Unix based systems have several readily available tools, for example PuTTY\citep{putty}, that achieve the same effect.  Requirement 3 recognises that some code will have to be available on each node that can load and invoke the application, but this should be as simple and easy as possible, and possibly independent of the application.  Requirement 4 means the application can be built and deployed on the network using different workstations and is not restricted to using a specific set of workstations.  Requirement 5 ensures the user can be confident the solution will work correctly.  Requirement 6 allows the user to build and deploy the application on different occasions using different host workstations and must only know the host’s IP-address.  Requirement 7 is not strictly necessary for correct functioning but does demonstrate the application is expected to terminate in an orderly manner.

\section{Background and Previous Work}\label{sec2}
ClusterBuilder uses the JCSP \citep{cspforjava} package, based on the occam \citep{inmos_ltd_1984} model that implements Hoare’s Communicating Sequential Processes \citep{Hoare-1978}.  The JCSP library provides the processes and channels required to build concurrent systems using Java.  The JCSP library has a formal proof of correctness for its operation\citep{welch_martin_2000Java}.The JCSP library includes a package JCSP.net2 that enables parallel processing over a TCP/IP network\citep{chalmers_2015}.  The great advantage of this approach is the process definitions are transparent as to whether communication is over internal or net channels.  The basic JCSP library has been further enhanced using several Groovy \citep{groovy_programming_language19} \citep{Konig_2015} classes that make programming systems even simpler, using GroovyJCSP \citep{kerridge_18_groovyJCSP}.  A previous library, Groovy Parallel Patterns\citep{kerridge_2016GPP}\citep{Kerridge-GPP-2021}, has been created that uses a similar DSL technology to build parallel applications on a single multi-core processor.  ClusterBuilder is an extension of that system to enable creation of network based parallel applications.  The Groovy Parallel Patterns Library has a formal proof of correctness using CSPm\citep{cspmref}  and FDR \citep{gibson-robinson_armstrong_boulgakov_roscoe_2014} and this approach has also been adopted for ClusterBuilder.

The concept of using workstation clusters is well known\citep{StoneandErcal2001}. The most common form of easily available cluster computing is known as a Beowulf cluster\citep{Beowulf} that describes the steps needed to create a simple Beowulf cluster using Ubuntu Linux.  This involves setting up specific files to hold node information, defining users, setting up a network file system to enable software distribution, setting up connections between the nodes using SSH and a process manager, typically Hydra\citep{hydra}.  It assumes the user is familiar with MPI\citep{MPI}\citep{OpenMP} and understands its intricacies.  This is a lot of technology to become familiar with and then confident in its use.  Conversely, workload management systems such as Condor\citep{HTCondor}\citep{UsingCondor} provides a means of controlling the resources of a workstation cluster using a batch control system.  The Condor system is able to use currently idle resource, on an as required basis, to satisfy the needs of jobs in its job queue.  Condor provides a management system that requires detailed knowledge of the requirements of a job, that new, or users with limited knowledge of parallel processing, may not have.

\section{The DSL Specification}\label{sec3}
Listing \ref{Listing1} shows the basic structure of an application specification with the cluster specific annotations \{1:2,4,6\}.  Initially, any constants used in the specification are defined.  

\begin{lstlisting}[caption={Outline DSL Specification for Cluster Builder},label=Listing1]
01.	... constants used in definition
02.	//@emit host-ip                         
03.	... emit process definition
04.	//@cluster Nclusters                    
05.	... cluster process definition
06.	//@collect                              
07.	... collect process definition
\end{lstlisting}

The \textit{//@emit} annotation specifies the IP-address of the host machine.  The processes that make up the object emit phase of the application are then defined.  The \textit{//@cluster} annotation specifies the number of workstations in the cluster and is followed by the process definitions to be created for each cluster.  The IP addresses for the cluster nodes are not required at this time.  Finally, the \textit{//@collect} annotation introduces the processes that make up the result collection phase of the application.  It is required that the emit and collect processes reside on the same host node.  In many applications having the emit and collect processes on the same node will not cause a reduction in performance as, by definition, most of the run-time will be consumed by the nodes undertaking the application processing.

\section{Network Utilisation During Application Loading}\label{sec4}
The network is structured and used in two completely different ways.  Initially, during application load time, the network is used in a specific manner over which the user has no control.  Only once the application has been loaded onto the cluster nodes does the network provide the interconnect required to undertake application processing.  The DSL creates a specific Host Node Loading application and the same Node Loading application on each node.  The DSL also creates a Host Process, comprising the Emit and Collect processes and a specific Node Process for each node.  The Node Process is created by the Host Node Loading application.

\begin{figure}[h!]
\centering
\includegraphics{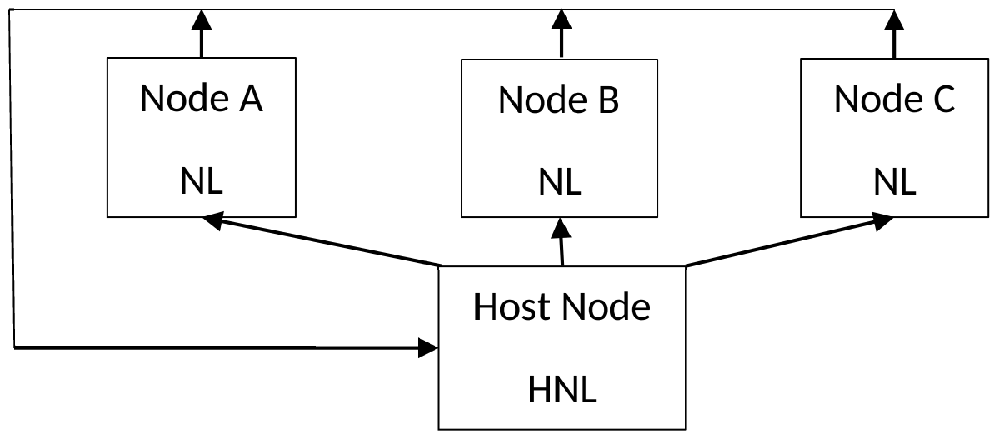}
\caption{Application Loading Network Architecture}
\label{Figure1}
\end{figure}

Figure \ref{Figure1}shows the network architecture used during application loading.  The Host Node Loading (HNL) application creates a many-to-one input channel to read messages from the nodes.  The user then must run an identical Node Loading (NL) application on each node, by transferring and running the same executable file from the host to each of the nodes, using a remote desktop connection.  The Node Loading application must be executed on the same number of nodes as specified in Nclusters \{1:4\}.  The NL then determines its IP-address and creates a one-to-one input channel that the HNL will use to output messages to the node.  The NL then sends its IP-address to the HNL.  Once the HNL has received the correct number of node IP-addresses, it creates the output channel to each of the nodes.  The HNL can now create the specific process required for each of the nodes that is obtained from running the ClusterBuilder application on the specification DSL.
The HNL can now send to each node the Node Process (NP) it is to execute, which contains enough information to be able to create the required application communication network.  The HNL then creates the Host Process (HP) which it loads into the host node.  Only when the HP is running can the application network be instantiated.  The JCSP.net2 package requires that an input channel end is created before the corresponding output end.  Thus, the initial part of the NPs and HP ensure this sequence is undertaken by sending synchronisation messages on the Application Loading Network shown in Figure 1.  The HP acts as the co-ordinating process.  Once all the application communication channels have been created the execution of the application can commence.  On termination of the application, the nodes return the timings for both loading and running the application.  The HP then combines these times with its equivalent times and displays the information.  Once this has completed all the nodes and the host processor are idle and all resources will have been reclaimed by the workstations’ operating systems.

\subsection{Class Loading}\label{sec4.1}
The JCSP.net2 package overcomes the challenge of loading class files by having the capability of accessing and dynamically loading them at run time.  This is achieved using code-loading channels which transform the class file into a serializable object for transmission and subsequently to de-serialize it when it is input into a reading process.  Thus, all the channels from the Host node in Figure \ref{Figure1} to the Nodes are code-loading channels.  The code-loading channel mechanism creates an additional channel from each Node back to the Host so that a Node can make a request for a class file if the required one is not available.  This means only the Host must have an instance of each class file, which it can then send to each Node after a request from the Node.  Class files can be loaded once the application has started running because the JCSP.net2 package ensures that only classes required by a Node are loaded.  The search for class files is not a generalised search of the complete network.  The search can only retrace through communication channels that have been declared as code-loading channels.  The search can pass through several nodes meaning there does not have to be a direct connection between a requesting node and the node that holds the class file.  Once the application terminates, the class files that have been loaded onto the Nodes are removed automatically, thereby ensuring file space is not consumed.

\section{An Example - The Mandelbrot Set}\label{sec5}
\begin{figure}[h!]
\centering
\includegraphics{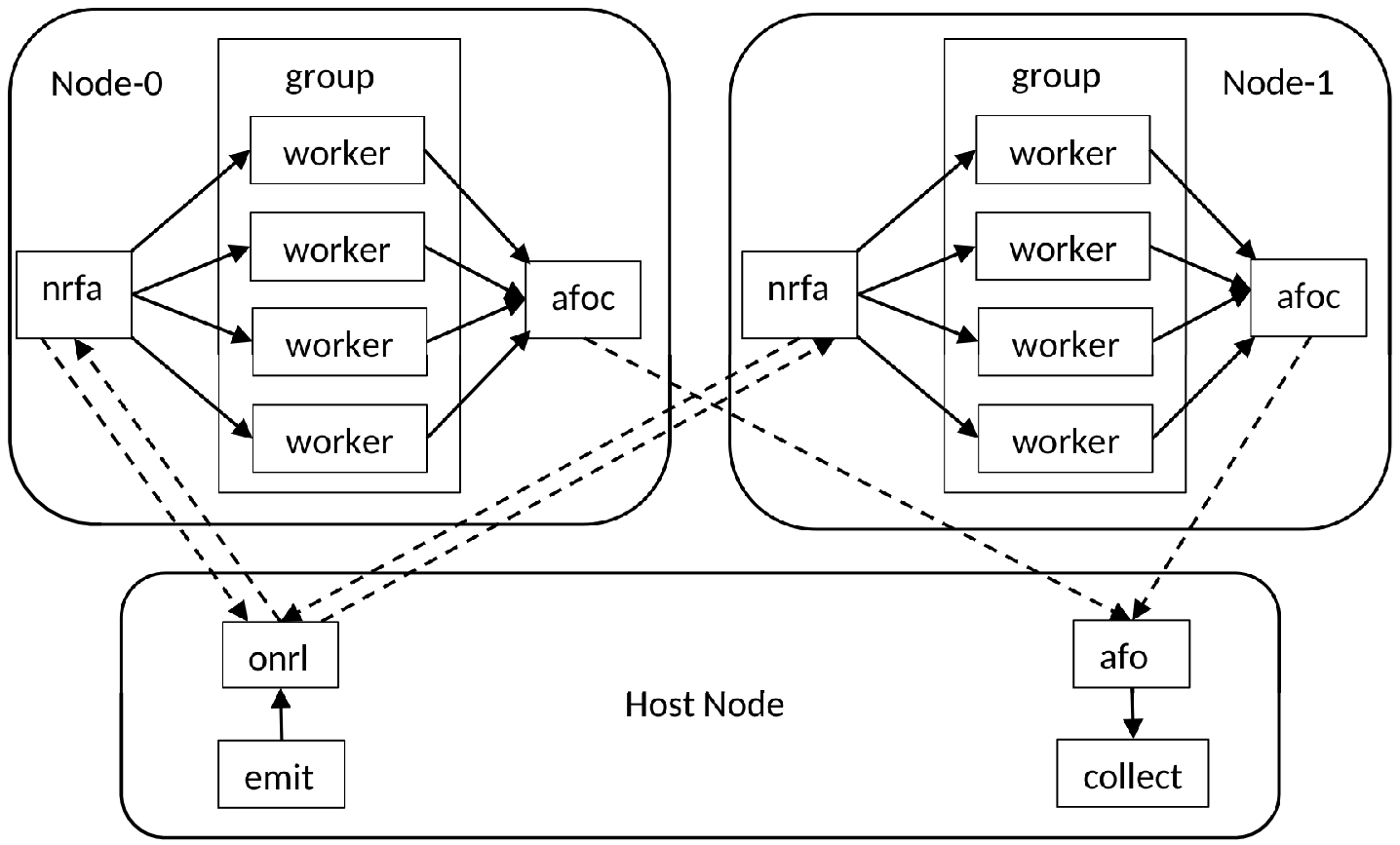}
\caption{Resultant Mandelbrot Process Network (Net Channels shown Dashed}
\label{Fig2}
\end{figure}

The Mandelbrot Set\citep{mandelbrot} is a well-known problem that can be solved using parallel techniques because it is a so-called embarrassingly parallel problem.  Listing \ref{Listing2} shows the DSL specification for a cluster-based solution and will produce the process network shown in Figure \ref{Fig2}.  It uses a single Host Node comprising the Emit and Collect processing and two numerical processing Nodes. 

\begin{lstlisting}[caption={Mandelbrot DSL Specification for 2 Nodes and a Host Workstation},label=Listing2]
01.	int cores = 4              // number of workers on each node
02.	int clusters = 2           // number of clusters
03.	int maxIterations = 1000   // escape value
04.	int width  = 5600          //double for more points
05.	
06.	//@emit 192.168.1.176
07.	def emitDetails = new DataDetails(
08.	                         dName: Mdata.getName(),
09.	                         dInitMethod: Mdata.initialiseClass,
10.	                         dInitData: [width, maxIterations],
11.	                         dCreateMethod: Mdata.createInstance )
12.	def emit = new Emit ( eDetails: emitDetails )
13.	def onrl = new OneNodeRequestedList()
14.	
15.	//@cluster clusters
16.	def nrfa = new NodeRequestingFanAny( destinations: cores )
17.	def group = new AnyGroupAny(
18.	                         workers: cores,
19.	                         function: Mdata.calculate )
20.	def afoc = new AnyFanOne( sources: cores )
21.	
22.	//@collect
23.	def resultDetails = new ResultDetails(
24.	                         rName: Mcollect.getName(),
25.	                         rInitMethod: Mcollect.init,
26.	                         rCollectMethod: Mcollect.collector,
27.	                         rFinaliseMethod: Mcollect.finalise )
28.	def afo = new AnyFanOne( sources: clusters )
29.	def collector = new Collect( rDetails: resultDetails )
\end{lstlisting}

The Mandelbrot space in the range x = -2.5 to 1.0 and y = -1.0 to 1.0 is subdivided into 3200 lines each of 5600 \{2:04\} points.  The \textit{emit} process \{2:12\} sends one complete line for processing by one of the Worker processes in either of the Nodes contained in the \textit{group} process \{2:17-19\}.  Finally, a processed line of data points is sent to the \textit{collector} process \{2:29\} where the image is reconstructed.  The class \textit{Mdata} used in \textit{emitDetails} contains \{2:7-11\} the definition of a line for a specified y-value, together with the methods used to initialise the class (\textit{initialiseClass}) and then create object instances one per line (\textit{createInstance}).  \textit{Mdata} also contains a method, called \textit{calculate}, that undertakes the determination of whether the point is in the Mandelbrot set (see Appendix B – Object Definitions). This method is invoked from one of the worker processes created by \textit{group} \{2:17-19\}.   The \textit{calculate} method uses an extant algorithm\citep{mandelbrot}, which uses \textit{maxIterations} \{2:03\} as the escape value.  As the \textit{Mdata} object is copied from the Host Node to the other nodes it must implement the \textit{Serializable} interface and the user must ensure that the data definition complies with serialization restrictions (no static or final properties can be serialized).  The remaining processes in the specification provide the communication between the processes and will be described shortly. 

Figure \ref{Fig2}shows the communication channels in more detail, with internal channels shown in solid lines.  ClusterBuilder creates all the required channels automatically.  It also creates the coding for the required HNL and NL applications and the HP and NP processes.
The emit process writes \textit{Mdata} objects to the \textit{onrl} process.  The \textit{onrl} and \textit{nrfa} processes form a client-server combination, with \textit{onrl} acting as the server.  The \textit{onrl} process reads an input from \textit{emit}.  It then waits for a request signal from any of the \textit{nrfa} client processes.  On reading such a signal, \textit{onrl} then responds by writing the input object to the requesting \textit{nrfa} process.  The \textit{nrfa} process then writes the object to any of the \textit{worker} processes that is currently idle.  Provided the \textit{onrl} server process responds to a client request in finite time and there is no loop in the client-server network then such a network is guaranteed to be deadlock and livelock free\citep{welch_justo_wilcock_1993}.  If all the \textit{worker} processes at a node are busy, then \textit{nrfa} acts as a one-place buffer until one of the \textit{worker} processes writes its processed object to \textit{afoc} and thus can read another input object for processing.  The \textit{nrfa} process cannot make a request for a new input object until it has written an object to a \textit{worker} process.  This ensures the \textit{onrl} process can never be blocked from writing an object to a node that has an idle \textit{worker} process.
The \textit{afoc} process can read an object from any of the \textit{worker} processes which it then writes, using a net channel to the \textit{afo} process that forms part of results processing.  Process \textit{afo} simply inputs any input object and writes it to the \textit{collect} process, where the results are collated.

\section{ClusterBuilder - Internals}\label{sec6}
Previously, in Section \ref{sec3}, reference was made to the creation of Host and Node Loading application (HNL, NL) and Host and Node Processes (HP, NP).  The ClusterBuilder contains proforma texts for each of these processes which are then modified according to the DSL specification.  The ClusterBuilder is an extension of the gppBuilder created for the Groovy Parallel Patterns (GPP) library\citep{kerridge_19a_GPPLibrary}.  The internal channels between processes are created by existing gppBuilder coding.  The ClusterBuilder is solely responsible for creating the net channels, together with the loading applications and node processes.  Some of the processes, for example in Figure 2, \textit{afoc} and \textit{afo} are specialisations of existing processes from the GPP Library which have a net output and input channel respectively which are created by ClusterBuilder.  ClusterBuilder is also responsible for net channel specific processes such as \textit{onrl} and \textit{nrfa}.

Net Channels have a specific address structure comprising node IP-address, port and channel number.  A computer in the cluster can support more than one logical node if they have different port numbers.  Thus, the load network, shown in Figure \ref{Figure1} uses port 2000 on all nodes and use channel number 1 for all interactions.  A net channel is solely defined by its net input channel address, thus the address of the input channel to the host is given by 192.168.1.176:2000/1 to which each of the nodes will output.  Each of the nodes will have a corresponding net channel input of the form 192.168.1.xxx:2000/1, where xxx is specific to the node.  Each of the nodes can determine their own IP-address and can send it to the host node because all they need is the host IP-address which is specified in the DSL specification \{2:6\}, provided the HNL is started before any of the NL processes.  It is a requirement of the JCSP.net2 package that the input end of a net channel is created before the output end is defined.  Such net channels are inherently a many to one connection and the underlying implementation ensures that inputs are processed in the order they are received.  A communication cannot start until a previous one is fully completed which includes sending an acknowledgement signal back to the sending process to ensure that all communications, either internal or net, are fully synchronised between sender and receiver.  To ensure that actions are undertaken in the correct order it is sometimes necessary to send synchronising messages between the nodes and the host to ensure that all nodes have completed a setup task before the host commences the next.

The DSL specification provides sufficient information for the HNL, together with the known structure of the processes that interact with net channels, to be able to construct node specific versions of the node process that take account of the net channel addresses that need to be used by each node.  In comparison, the NL process initially executed by each node is totally application independent such that once a node has an instance of the NL executable it can be used to load and execute any application created using the DSL.  The first part of an application node process is in fact concerned with setting up the application net channels.  Only when this is done can the HNL load and invoke the application emit and collect processes.

\subsection{IDE Integration}\label{sec6.1}
The DSL specification is contained in a file with a \textit{.cgpp} suffix.  Within the IDE the file association \textit{cgpp} is linked to Groovy so that syntax checking of the specification can be undertaken.  The ClusterBuilder application is applied to the file which creates four output files with the .groovy suffix.  These take the form of the DSL file name followed by HostLoader, HostProcess, NodeLoader and NodeProcess.  Of these only the app\_name\_NodeLoader.groovy file needs to be made into an executable.  In most IDE this can be automated by the production of a jar type artefact that can be invoked using the java -jar file command line.  This jar-file needs to be copied to all the other computers in the cluster upon which the application is to run, most easily by using a remote desktop connection.  The HostLoader groovy file can be executed directly from the IDE, after which the jar file at each node can be invoked.  The loading of the processes then follows automatically using the Node and Host Process files.  All class files are loaded automatically from the host computer to the nodes without any intervention from the user.  On termination the nodes send some timing data to the host before terminating and recovering all resources.  The host prints out any required results and all the timing data before itself finishing.  
In some situations, it may be beneficial to run the host from the command line.  This can be achieved by creating a jar of the HostLoader process.  This will contain all the required class files to run the application.  Care must be taken to ensure the host IP-address is correctly specified and the number of worker nodes is correct as the ClusterBuilder encodes these into the Host Loader process.  The NodeLoader will also need to match the created host as it also encodes the host IP-address.
The operation and testing of a system can be conducted on a single host node before using multiple nodes.  The number of clusters is set to 1, the host loader process is then run followed by node loader process directly from the IDE.  The application will be loaded and executed directly using the IDE thereby enabling confidence building.  This is possible because the application network uses a different port from the process loading phase.  Obviously, the performance will be not as fast as using multiple nodes, but it is now known the application will load correctly.

\section{Formal Verification of the Cluster Architecture}\label{sec7}

The processes included in the GPP Library have been formally shown to be correct \citep{Kerridge-GPP-2021} using CSPm\citep{cspmref} specifications and the FDR\citep{gibson-robinson_armstrong_boulgakov_roscoe_2014} checking tool.  The client-server protocol described earlier, which is based on an original idea of Brinch Hansen\citep{hansen_1973}, also has a formal proof\citep{welch_martin_2000CSP}.  The CSPm specification shown in Listing \ref{Listing3} contains the definition of all the \textit{datatypes}, \textit{objects}, \textit{channel}s and \textit{process}es used in the architecture.  

\begin{lstlisting}[caption={CSPm Specification of the Cluster Architecture},label=Listing3]
01.	datatype objects = A | B | C | D | E | UT  
02.	datatype signal = S
03.	N = 2               	     // number of cluster nodes
04.	channel a: objects           // channels connecting the processes
05.	channel b: {0..N-1}.signal   // indicating the object types that
06.	channel c: {0..N-1}.objects  // form the events on the channel  
07.	channel d: {0..N-1}.objects   
08.	channel e: {0..N-1}.objects   
09.	channel f: objects
10.	channel finished: Bool       // used during system refinement
11.	a_A = {|a|}   // the alphabet (events) associated with each channel
12.	a_B = {|b|}
13.	a_C = {|c|}
14.	a_D = {|d|}
15.	a_E = {|e|}
16.	a_F = {|f|}
17.	create(A) = B      // the specification of objects as they
18.	create(B) = C      // are created and emitted from the
19.	create(C) = D      // Emit process
20.	create(D) = E 
21.	create(E) = UT     // the final terminating Universal Terminator
22.	Emit(o) = a!o -> if o == UT then SKIP 
23.	                            else Emit(create(o))
24.	Server() = a?o -> if o == UT then Server_End(0) 
25.	                             else  Server_Choice(o) 
26.	Service(i, o) = b?i.S -> c!i.o -> Server()
27.	Server_Choice(o) = [] x : {0..N-1} @ Service(x, o)
28.	Server_End(y) = b?y.S -> c!y.UT -> if y == N then SKIP 
29.	                                             else Server_End(y+1)
30.	Client(i) = b!i.S -> c?i.o -> if o == UT then (d!i.UT -> SKIP) 
31.	                                         else (d!i.o -> Client(i))
32.	a_Client(x) = {|b.x, c.x, d.x|}	// client alphabet
33.	Clients() = || x : {0..N-1} @ [a_Client(x)] Client(x) // parallel
34.	a_CS = {|b, c|}			// client-server combined alphabet
35.	Worker(i) = d?i.o -> if o == UT then (e!i.UT -> SKIP) 
36.	                                else (e!i.o -> Worker(i))
37.	a_W(x) =  {|d.x, e.x|}  // the alphabet used by each Worker(x)
38.	Workers() =  || x : {0..N-1} @  [a_W(x)] Worker(x) // parallel 
39.	Reduce(i) = e?i.o -> if o == UT then (Reduce_End(i, (i+1)%N) ) 
40.	                                else ( f ! o -> Reduce(i))
41.	Reduce_End(s, n) = if s==n then f ! UT -> SKIP 
42.	                      else e?n.o -> 
43.	                          if o == UT then ( f!o -> Reduce_End(s, n) ) 
44.	                                     else Reduce_End(s, (n+1)%N ) 
45.	Reducer() = [] x: {0..N-1} @ Reduce(x)    // replicated choice
46.	Collect() = f?o -> if o == UT then Collect_End() 
47.	                              else Collect()
48.	Collect_End() = finished ! True -> Collect_End()
49.	
50.	System = ((((Emit(A) [| a_A |] Server()) [|a_CS|] Clients()) 
51.	      [| a_D |]  Workers() ) [| a_E |] Reducer() ) [| a_F|] Collect()
52.	TestSystem = finished!True -> TestSystem   
53.	assert (System \ {|a, b, c, d, e, f|}) [T= TestSystem 
54.	assert (System \ {|a, b, c, d, e, f|}) [F= TestSystem
55.	assert (System \ {|a, b, c, d, e, f|}) [FD= TestSystem
56.	assert System :[deadlock free]
57.	assert System :[divergence free]
58.	assert System :[deterministic]

\end{lstlisting}

\begin{figure}[h!]
\centering
\includegraphics{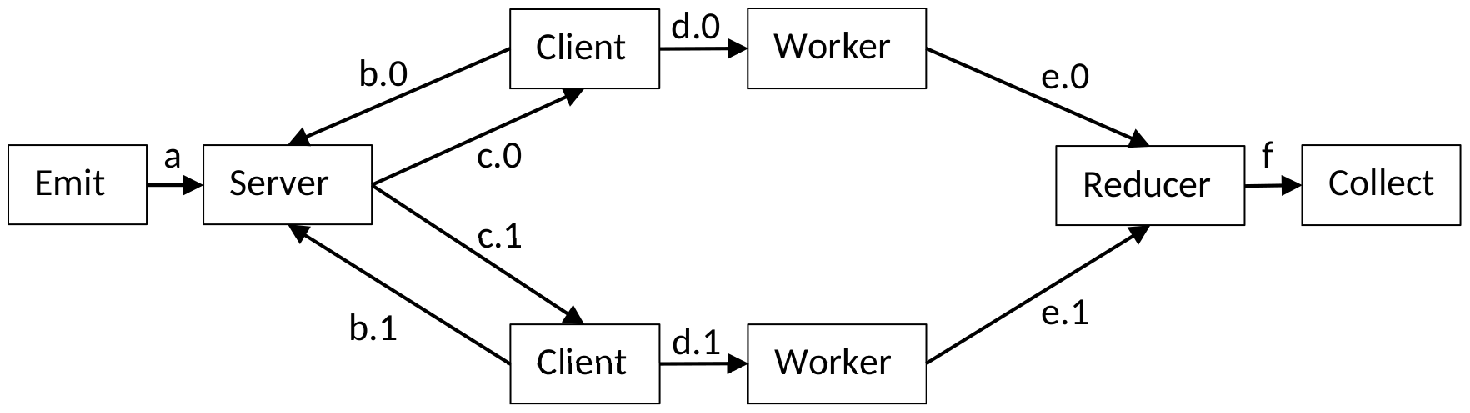}
\caption{Channel and Process Structure Used in Formal Proof}
\label{Fig3}
\end{figure}

Each of the processes in Listing \ref{Listing2} have their own model in the specification.  The state space that must be investigated is reduced by emitting only five objects \{3:1\} into the network together with \textit{UT}, a termination object.  The \textit{signal} object \{3:2\} is used only by the \textit{Client} process to indicate that it needs a new input from the \textit{Server}.  The channel and process structure used by the specification is shown in Figure \ref{Fig3}.

In terms of the specification, \textit{onrl} is modelled by \textit{Server}, \textit{nrfa} by \textit{Client}, the group of \textit{workers} by \textit{Worker} and \textit{Reducer} combines the effect of the \textit{afoc} and \textit{afo} processes.  Each of the channels, \textit{a} to \textit{f} \{3:4-10\} is defined and the object types they communicate, and where necessary indexed by the number of clusters N \{3:3\}.  A channel follows the CSP definition\cite{Hoare-1978} which is unidirectional, synchronised and unbuffered means of writing (!) from one process to a single reading (?) process.  The channel \textit{finished} is used in the model (\textit{TestSystem} \{3:52\}) against which the \textit{System} is compared to check the deadlock and livelock status.  The basis of the assertion checking \{3:53:58\} is that \textit{System} \{3:50-51\} can only behave the same as \textit{TestSystem} provided the model shown in Figure 3 and Listing 2 works correctly.

The alphabets \{3:11-16\} specify the set of events that each channel can recognise.  The \textit{create} functions \{3:17-21\} are used by the \textit{Emit} \{3:22-23\} process to write the sequence of objects \textit{A} to \textit{E} into the network.  It then writes \textit{UT} into the network and then behaves as SKIP.  The process SKIP indicates a process has terminated correctly.  The object \textit{UT} will be passed through all the processes, in turn, causing them to successfully terminate, except \textit{Collect} \{3:46-48\}, which repeatedly outputs True on the \textit{finished} channel, the same as the\textit{ TestSystem}.  Once FDR detects this situation it can undertake the required assertion evaluation.
The definition of the \textit{Server} \{3:24-19\} is the most complex comprising four functions.  \textit{Server()} reads an object from channel \textit{a} and determines if it is \textit{UT}, in which case \textit{Server\_End(0)} is called.  This will cause \textit{UT} to be written to each of the \textit{Client} processes in order.  If the object is one of \textit{A} to \textit{E} then a replicated non-deterministic choice is invoked by \textit{Server\_Choice(o)}.  The \textit{Service(i.o)} function undertakes a \textit{signal} input from one of the \textit{Client b} channels and then writes a data object on the \textit{c} channel with the same index value.

The \textit{Clients()} processes are a parallel replication of \textit{Client()} processes \{3:30-34\}.  A \textit{Client} writes a \textit{signal} to the \textit{Server} using its \textit{b} channel which will only be read once the \textit{Server} has read an input from \textit{Emit}.  It then reads an input on its \textit{c} channel and then writes the value to its \textit{d} channel taking account of the \textit{UT} object appropriately.

The \textit{Workers} process is also a replicated parallel of \textit{Worker} processes \{3:35:38\}, which input from their \textit{d} channel and output the object on the \textit{e} channel also taking account of the \textit{UT} object correctly.  The \textit{Worker} in the model undertakes no function but this has no effect on the modelling of the communication structures.
The \textit{Reducer} process \{3:39-45\} uses a non-deterministic replicated choice to read objects from any of the \textit{e} channels.  It detects the input of a \textit{UT} object on one of the \textit{e} channels and ensures that any non-UT objects are processed before reading \textit{UT} objects from the other \textit{e} channels and then finally outputs a single \textit{UT} object to the \textit{f} channel. The \textit{Collect} process \{3:46-48\} inputs objects from the \textit{f} channel until a \textit{UT} is read at which point the process outputs True to the \textit{finished} channel.

FDR takes the specification and assesses the assertions which determine the correctness of the specification.  A user can also use the \textit{:probe System} at the FDR prompt and can step through the specification choosing available events.  For even such a small specification it soon becomes obvious the large number of possible event orderings, known as Traces, that can occur.
\section{Performance Evaluation}\label{sec8}
The performance of the Mandelbrot application is evaluated in two ways.  First on a single processor with 16 cores to evaluate the effect of using a single powerful machine and then on a cluster of much less powerful machines to evaluate the performance improvement that can be achieved using a workstation cluster.  In both evaluations the same data objects were used, without alteration (see Appendix B).  The number of points per line was 5600 on the 3200 lines, giving a total of 17.92 million points of which just over 14 million were white.  The total number of iterations undertaken in doing all the calculations was 3,962 million using an escape value of 1000.  The colour of each point was stored in an integer array and the co-ordinates of each point were stored in a two-dimensional array of doubles.  The results were identical regardless of the processing resource used.
\subsection{Single Processor Performance}\label{sec8.1}
An X-series Intel i9-7960X overclocked at 4.40Ghz with a 22MB cache and 64GB RAM was used for the evaluation.  The processor also has hyper-threading to a maximum of 32 hyper-threads.  The multi-core parallel architecture uses the same \textit{Emit} and \textit{Collect} and \textit{Group} of worker processes, varying the number of workers between 1 and 32.  Table \ref{Table1} shows the performance achieved with varying number of \textit{Worker} processes.

\begin{table}[]
\centering
\begin{tabular}{|c|c|c|c|}
\hline
Workers & Time msecs & Speedup & Efficiency \% \\ \hline
1       & 882963     &         &               \\ \hline
2       & 447175     & 1.97    & 98.73         \\ \hline
4       & 221139     & 3.99    & 99.82         \\ \hline
8       & 115890     & 7.62    & 95.24         \\ \hline
12      & 89970      & 9.81    & 81.78         \\ \hline
16      & 90173      & 9.79    & 61.20         \\ \hline
20      & 87215      & 10.12   & 50.62         \\ \hline
28      & 94418      & 9.35    & 33.40         \\ \hline
32      & 100232     & 8.81    & 27.53         \\ \hline
\end{tabular}
\caption{Single Processor Performance}
\label{Table1}
\end{table}

The Standard Deviation of the Time was less than 7{\%} of the mean in all cases bar, 1 Worker - 8.6\% and 32 Workers - 9.2\%.  Speedup is a measure of the reduction in time compared to the single Worker version (T1/Tn  n = 2 to 32).  The Efficiency (Speedup / Workers) measures, as a percentage, how effective the use of the multiple Worker processes was.  Ideally, the Speedup should be close to the number of Workers, which for the cases 2, 4 and 8 is excellent with an Efficiency of better than 95\%.  In the case of 16 Workers, the number of available cores, Speedup is only just better than with 8 cores and the Efficiency is correspondingly worse.  The network comprises \textit{Emit}, \textit{Collect} and the \textit{Worker} processes plus two further processes that distribute the work packets and then combine them together before final collection.  This is more than the available processor resource, but the additional processes will be doing very little work in comparison to the Worker processes.  The version with 28 Workers was run to determine the effect of using all 32 hyper-threads.  As can be seen it is better than using 32 hyper-threads but still worse than using 20 Workers.  Thus, even though the problem is embarrassingly parallel the processor configuration means that performance reduces as the number of workers increases.  This can possibly be accounted for by the interaction and contention between each of the cores accessing the single processor cache memory.
\subsection{Cluster Performance}\label{sec8.2}
This experiment was carried out on a network of Intel i7-8700 computers running at 3.2GHz with a 12MB cache.  Each processor has 6 cores and a further 6 hyper-threads and 16GB RAM.  The Nodes were organised as 4 \textit{workers} plus the \textit{nrfa} and \textit{afoc} processes (see Figure \ref{Fig2}), thereby using all the available cores.  One workstation acted as the Host Node with additional nodes being added, thereby enabling comparison with the single multi-core processor.  Based on the previous experiments, it was decided not to make use of hyper-threads.  The nodes were connected by a 1GB ethernet and the results are shown in Table \ref{Table2}.

\begin{table}[]
\centering
\begin{tabular}{|c|c|c|c|c|}
\hline
Nodes & Cores & Time*  & Speedup  & Efficiency \\ \hline
0     & 4     & 243425 &          &            \\ \hline
1     & 4     & 230771 & 1.054837 & 105.5\%    \\ \hline
2     & 8     & 120912 & 2.013251 & 100.7\%    \\ \hline
3     & 12    & 82237  & 2.960049 & 98.7\%     \\ \hline
4     & 16    & 84301  & 2.887584 & 72.2\%     \\ \hline
5     & 20    & 75122  & 3.240418 & 64.8\%     \\ \hline
\end{tabular}
\caption{Cluster Performance Evaluation \\ * Time in milliseconds averaged over 10 runs}
\label{Table2}
\end{table}

The zero nodes time refers to the base case where both the Host and the single node processes were running on the same machine.  For the cases using 1 and 2 additional nodes the speedup compared to the base case is super-linear and thereafter there is a drop off in speedup.  However, the comparative efficiency for the same number of cores as seen in Table 1 is better in the Cluster version.

Table \ref{Table3} shows the mean elapsed time for the multi-core and the cluster versions using the number of worker cores as the basis for comparison.  The cluster machines processor frequency was 37.5\% slower than the multi-core machine.

\begin{table}[]
\centering
\begin{tabular}{|c|c|c|c|}
\hline
Worker Cores & Multi-core Tm & Cluster Tc & Difference (Tc-Tm)/Tc \\ \hline
4            & 221139             & 230771          & 4.2\%                 \\ \hline
8            & 115890             & 120912          & 4.2\%                 \\ \hline
12           & 89970              & 82237           & -9.4\%                \\ \hline
16           & 90173              & 84301           & -7.0\%                \\ \hline
20           & 87215              & 75122           & -16.1\%               \\ \hline
\end{tabular}
\caption{Performance Comparison Between Multi-core and Cluster Solutions \\ 
All Times in milliseconds and averaged over 10 runs}
\label{Table3}
\end{table}

For 4 and 8 worker cores the multi-core machine (16 cores) was faster by a small percentage but once the application was loaded onto a cluster of smaller (6 cores) and slower machines the cluster gave the better performance in terms of total run time.  In all cases the application load time, once all the nodes had sent their IP-address to the host, was less than 1\% of the total application run time, more significantly, the increase in load time was linear in the number of nodes, 132.5 +/- 2.5 milliseconds, for this application.
\section{Conclusions and Future Work}\label{sec9}
The experiments have demonstrated that a parallelisable application can be run on a cluster of small multi-core machines with better scalability than running on a single faster larger multi-core machine.  The problem then is one of allocating and loading code over the cluster.  The ClusterBuilder application has satisfied the requirements outlined in Section \ref{sec1.1} and achieves the goal of taking a sequential solution and parallelising it both on a single and a cluster of machines.  The basic sequential solution must be augmented by a small number of relatively simple methods (see Appendix B) to enable interaction with the supporting parallel library.  The solution is fully integrated with existing IDEs, Intellij\citep{intellij}, in this case. Furthermore, the generated solution has a formal proof of correctness.  

The additional user knowledge required is very limited, Listing 2 and the class definitions given in Appendix B, present the total programming required to achieve a cluster based parallel solution.  The major step is appreciating the route to parallelisation of an algorithm is splitting up the data in such a way as to allow secure parallel access.  In this case, splitting the Mandelbrot space into lines and processing each line by itself, rather than in a sequential solution processing the whole space as a single entity.

Future work will concentrate on two aspects.  First, the ability to create a sequence of different algorithms, each on their own cluster and secondly, ensuring that certain nodes in the network can be fixed so that node specific access, say to data storage or by means of a network file system, is enabled.

%%\subsection{Bibliography}
%%\nocite{*}% Show all bib entries - both cited and uncited; comment this line to view only cited bib entries;

\bibliographystyle{unsrtnat}
\bibliography{references}

\appendix
\section{Software Availability}
The primary download for all the demonstration software used in this paper is \url{https://github.com/JonKerridge/ClusterDemos}.
Its build file will download all the other libraries used by the library including the groovy\_parallel\_patterns library and its associated gppClusterBuilder program.  Once a reader has decided they wish to delve further they may want to look at all the libraries used in the demonstration system.  The README in the repository gives the location of the repositories used and the actual dependency of the library.
The library software also requires the JavaFX capability, used by a visualisation capability, but this is downloaded as part of the build file for the Groovy Parallel Patterns library.  The version used is version 11 and thus an environment using the groovy\_parallel\_patterns library must use Java JDK 11 and Groovy.3.  

To download Packages from the Github Package Repository a user must provide a personal Access Token, see 
\url{https://docs.github.com/en/github/authenticating-to-github/creating-a-personal-access-token}
This token must be made available in a file called gradle.properties. This file should not be saved in a repository as the Personal Access Token acts in the same way as a password. The build.gradle file contains a mechanism that accesses the gradle.properties file from a local folder (C:/Github/gradle.properties) that can be accessed by the Gradle build mechanism. It should contain the two lines.

\tt{gpr.user=userName \\}
\tt{gpr.key=userPersonalAccessToken}

\section{Object Definitions}

\begin{lstlisting}[caption={Mdata and Mcollect definitions}]
01.	class Mdata extends groovyParallelPatterns.DataClass{
02.	  int []colour       // array of colour values for this line
03.	  double [][] line   // array of [x,y] values for this line
04.	  double ly          // y value for this line
05.	  int escapeValue    // Mandelbrot maximum iterations before escape
06.	  long totalIterations // total iterations per line 
07.	  int WHITE = 1
08.	  int BLACK = 0
09.	  double minX = -2.5   // range over which calcualtions undertaken
10.	  double minY = 1.0
11.	  double rangeX = 3.5
12.	  double rangeY = 2.0
13.	  static String initialiseClass = "initClass"
14.	  static String createInstance = "createInstance"
15.	  static String calculate = "calculateColour"
16.	  static int lineY = 0  // values used by createInstance
17.	  static int heightPoints, widthPoints, maxIterations
18.	  static double delta
19.	  int initClass ( List d){
20.	    widthPoints = (int) d[0]
21.	    maxIterations = (int) d[1]
22.	    delta = rangeX / ((double) widthPoints)
23.	    heightPoints = (int) (rangeY / delta )
24.	    return completedOK
25.	  }
26.	  int createInstance (List d) {
27.	    if (lineY == heightPoints) return normalTermination
28.	    colour = new int[widthPoints]     // instance variables
29.	    line = new double[widthPoints][2]
30.	    escapeValue = maxIterations
31.	    totalIterations = 0
32.	    ly = lineY * delta //y value for this line
33.	    0.upto(widthPoints-1){ int w ->
34.	      line[w][0] = minX + ( w * delta)
35.	      line[w][1] = minY - ly
36.	    }
37.	    lineY = lineY + 1
38.	    return normalContinuation
39.	  }
40.	// based on algorithm at https://en.wikipedia.org/wiki/Mandelbrot_set 
41.	 int calculateColour (List d) {
42.	    int width = colour.size()
43.	    0.upto(width-1){ int w->
44.	      double xl = 0.0, yl = 0.0, xtemp = 0.0
45.	      int iterations = 0
46.	      while (((xl * xl)+(yl * yl) < 4) && iterations < escapeValue) {
47.	        xtemp = (xl * xl) - (yl * yl) + line[w][0]
48.	        yl = (2 * xl * yl) + line[w][1]
49.	        xl = xtemp
50.	        iterations = iterations + 1
51.	      }
52.	      totalIterations += iterations
53.	      colour[w] = (iterations < escapeValue) ? WHITE : BLACK
54.	    }
55.	    return completedOK
56.	  }
57.	}
58.	
59.	class Mcollect extends groovyParallelPatterns.DataClass {
60.	  int blackCount = 0  // counts for points and black and white
61.	  int whiteCount = 0
62.	  int points = 0
63.	  long totalIters = 0  // sums total iterations over all lines
64.	  static String init = "initClass"
65.	  static String collector = "collector"
66.	  static String finalise = "finalise"
67.	  int initClass ( List d){
68.	    return completedOK
69.	  }
70.	  int finalise( List d){
71.	    println "$points, $whiteCount, $blackCount, $totalIters "
72.	    return completedOK
73.	  }
74.	  int collector(Mdata ml){
75.	    int width = ml.colour.size()
76.	    0.upto(width-1){ int w->
77.	      points = points + 1
78.	      if (ml.colour[w] == ml.WHITE) whiteCount = whiteCount + 1
79.	      else blackCount = blackCount + 1
80.	    }
81.	    totalIters += ml.totalIterations
82.	    return completedOK
83.	  }
84.	}

\end{lstlisting}
\end{document}